\documentclass{article}

\usepackage{graphicx, verbatim, url, epigraph, enumerate}
\usepackage[utf8x]{inputenc} 

\begin{document}

\title{Complex Beauty}

\author{Massimo Franceschet\footnote{Massimo Franceschet is also an unpretentious contemporary dancer and apprentice generative artist.} \\
Department of Mathematics and Computer Science \\
University of Udine \\
\url{massimo.franceschet@uniud.it}
}
\maketitle

\epigraph{kono michi ya \\ yuku hito nashi ni \\ aki no kure \\
All along this road \\ not a single soul \\ only autumn evening comes}
{Bash\={o} Matsuo}

\begin{abstract}
Complex systems and their underlying convoluted networks are ubiquitous, all we need is an eye for them. They pose problems of organized complexity which cannot be approached with a reductionist method. Complexity science and its emergent sister network science both come to grips with the inherent complexity of complex systems with an holistic strategy. The relevance of complexity, however, transcends the sciences. Complex systems and networks are the focal point of a philosophical, cultural and artistic turn of our tightly interrelated and interdependent postmodern society. Here I take a different, aesthetic perspective on complexity. I argue that complex systems can be beautiful and can the object of \textit{artification} - the neologism refers to processes in which something that is not regarded as art in the traditional sense of the word is changed into art. Complex systems and networks are powerful sources of inspiration for the generative designer, for the artful data visualizer, as well as for the traditional artist. I finally discuss the benefits of a cross-fertilization between science and art. 
\end{abstract}


\section{Art as therapy}

The idea for this contribution comes from a reflection that, on the surface, may appear paradoxical: in a period of scarcity of resources, ideas, and imagination, as the one we are experiencing, we need to invest our time and energy in art, committing ourselves to creating and sustaining \textit{beauty} so that it can be enjoyed freely by people, and in the process reconnect with our most vulnerable self, without which no meaningful art would see birth. To be sure, this thought emerged after having tasted \textit{Art as Therapy} \cite{deBA13}. Philosopher Alain de Botton teams up with art historian John Armstrong and expose, in an engaging and lively way, a basic proposition: far more than mere aesthetic gratification, art is a \textit{therapeutic tool} that can improve the quality of our lives. Benefiting from inspiring, cogent pictorial artworks, the authors argue that art, as a tool, has seven neat psychological functions that might be useful to allay as many human frailties: art works can help us to remember what matters; they also lend us hope; they dignify sorrow; they expand our horizons; they help us to understand ourselves; they rebalance us; and lastly they make us appreciate the familiar anew: 

\begin{quote}
Art is one resource that can lead us back to a more accurate assessment of what is valuable by working against habit and inviting us to recalibrate what we admire or love. (...) Art can do the opposite of glamorizing the unattainable; it can reawaken us to the genuine merit of life as we're forced to lead it.
\end{quote}

Everyone can give a contribution to this thesis, not only the genuine artist. The scientist as well. In the following I will focus on complex systems, my principal research interest. I will argue that complex systems, besides being an established tool to investigate reality, are extremely alluring processes generating beautiful networks. As unstable, soft blend of order and disorder, wildly distributed in technology, information, society and nature, complex systems are a precious implement for the generative artist, a new inspiring source for the traditional artist, as well as a varicolored data set for the artful information visualizer.
      
\section{The ubiquity of complex systems}  

In 1948 american scientist Warren Weaver wrote a much discerning article entitled \textit{Science and Complexity} \cite{W48}, anticipating the advent of a new science of networks devoted to the investigations of complex systems. Weaver spoke of `\textit{problems of organized complexity}'. Such problems ``involve dealing simultaneously with a sizable number of factors which are interrelated into an organic whole.'' According to Weaver, the solution of such problems requires science to make a great advance, exploiting a mixed-team (interdisciplinary) approach: ``It was found, in spite of the modern tendencies toward intense scientific specialization, that members of such diverse groups could work together and could form a unit which was much greater than the mere sum of its parts. It was shown that these groups could tackle certain problems of organized complexity, and get useful answers.''

A comprehensive description of the characteristics of complex systems is given by philosopher and complexity researcher Paul Cilliers in a book that draws a fascinating connection between complexity and post-modernism \cite{C98}. \textit{Complex systems} consist of a large, interacting number of actors. Interactions are dynamic (they change with time), fairly rich (actors typically influence quite a few other ones), mostly short-range (information, or whatever else might circulate through relationships, is received and spread primarily from immediate neighbors), non-linear (small causes can have large effects and vice versa), and non-hierarchical (there are feedback loops in relationships). Actors are self-organizing (there exist no central authority) and ignorant of the behaviour of the system as a whole (they have local information only). Furthermore, the system is open (interacting with the environment), operates under conditions far from equilibrium (it is kept alive by a constant flow of information), and has a history (the past is co-responsible for the present behaviour). Complex systems are widespread in nature, society, information and technology; a few examples include: the human brain, the metabolic system, the natural language, ecosystems and the biosphere, the academic publication system, linked information systems like the Web and Wikipedia, online social networking services such as Twitter, LinkedIn and ResearchGate, the economic system, the Internet and power grids. 

The difficulty with complex systems is that they are \textit{complex}, not merely \textit{complicated}. The very peculiarity of a complex system lies in the relationships among its parts. Such an inseparable coupling makes the system more than the mere juxtaposition of its parts, hence the system as a whole cannot be fully understood simply by analyzing its components. Consider the Web, for instance. The content of a Web page tells us only half of the story; it is useful to define the relevance of a page with respect to a user's information need. The hyperlinks between pages complete the picture: they contain the precious information that can be used to gauge the importance of a page with algorithms such as PageRank \cite{F11-CACM}. Similarly, the scholarly papers we write are of incommensurable value; on the other hand, bibliographic citations among them are also important to measure their impact \cite{WBB10}.

Reductionism - an analytical method that analyses something complex by dividing it into manageable parts which can be investigated separately and then by putting the parts together again - is not a useful strategy with complex systems. As Cilliers says: ``In `cutting up' a system, the analytical method destroys what it seeks to understand'' \cite{C98}. On the other hand, \textit{holism} - which believes that the whole is ultimately irreducible - is a more viable approach to the understanding of complex systems.\footnote{To be sure, the general principle of holism can be tracked back to Aristotle, who used it to respond to a famous paradox put forward by Zeno (a distance is first and foremost an irreducible whole). A more recent embodiment of the concept lies in Gestalt psychology. Gestaltists assert that the brain is holistic: human experiences cannot be derived from the summation of perceptual elements, because they are ultimately irreducible \cite{G99}.} Complex systems pose real problems of organized complexity, as Weaver anticipated, and that demands new ways of thinking.

A feasible, although incomplete, approach to the inherent complexity of complex systems is \textit{network science} - the holistic analysis of real complex systems through the study of the network that wires their components  \cite{B02,N10}. It is worth noticing that a network is a simplified, partial model of a complex system: it captures only the structure of relationships among actors, which is, nevertheless, the most valuable and tasty aspect of complex systems. 

The first tangible contribution of network science has been  the collection of network data: the identification, construction, storage, and distribution of a differentiated database of possibly very large real networks. These networks underlie complex systems present in many different contexts including technological networks (Internet, telephone networks, power grids, transportation networks, and distribution networks), information networks (the Web, academic and legal citation networks, patent networks, peer-to-peer networks, and recommender networks), social networks (friendship and acquaintance networks, collaborations of scientists, movie actors, and musicians, sexual contact networks and dating patterns, criminal networks, and social networks of animals), as well as biological networks (metabolic networks, protein-protein interaction networks, genetic regulatory networks, neural networks, and ecological networks). 

Network scientists study methods and realize tools to analyze such a rich repository of real graphs. Some of these methods are new (for instance, algorithms for community detection and characterization of networks according to their node degree distribution), other are indeed borrowed from graph theory, bibliometrics, sociometry and even econometrics. Network science addresses questions at three levels of granularity \cite{BE04}: node-level analysis, where methods to identify the most central nodes of the network are investigated,  group-level analysis, that involves techniques for defining and finding cohesive groups of nodes in the network, and network-level analysis, that focuses on topological properties of networks as a whole as well as on theoretical models generating empirical networks with certain properties.   

Nevertheless, complex systems and networks are not just \textit{useful}; they are also \textit{beautiful}: networks can be artworks. At least, this is the claim of the following section. 
   
\section{The beauty of complex systems}  
    
There exists a general consensus in aesthetics - the philosophical study of art, beauty and taste - that beauty lies at intersection of \textit{order} and \textit{disorder}. The perfect order is tedious and therefore not attractive. The chaos is incomprehensible to our brain and therefore is equally unappetizing. When we depart from order without resulting in complete chaos, maintaining an unstable balance between regularity and mess, often we get a result that surprises and thrills, so that we may define it beautiful. Consider a performance of contemporary dance. Each involved dancer typically follows a specific choreography, determined a priori by the choreographer. On the other hand, each dancer interprets the choreography according to their inclinations, history, and mood. Not infrequently, it is also left room for improvisation. These elements - interpretation and improvisation - add a disorderly contribution to the choreographed, pre-given movements. It follows that every staging is the same but also subtly different from the others; it is partially unpredictable. 

Architect Richard Padovan describes order and complexity as twin poles of the same phenomenon. Neither can exist without the other - order needs complexity to become manifest, complexity needs order to become intelligible - and aesthetic value is a measure of both. He beautifully expresses this concept with the following words: ``Delight lies somewhere between boredom and confusion. If monotony makes it difficult to attend, a surfeit of novelty will overload the system and cause us to give up; we are not tempted to analyze the crazy pavement.'' \cite{P02}. 

We argue that complex systems live at the edge of chaos, at the intersection of order and disorder. If we look complex systems at the micro level of actors, they appear relatively simple and regular systems. Individual actors operate in a rather elementary way, typically following few plain rules, paying attention to the behaviour of their local neighbors only. Such a local simplicity, multiplied by the sheer number of actors that compose the system, and, moreover, amplified through the convoluted structure of relationships among actors, produce an unexpected, yet organized, global complexity. A simple rule set at a low level creates organized complexity at a higher level. 

Let me give a couple of examples. In a bird flock, according to the simplest model \cite{R87},  each individual bird maneuvers based on the positions and velocities of its nearby flock mates following three simple steering behaviors: separation (steer to avoid crowding local flock mates), alignment (steer towards the average heading of local flock mates), and cohesion (steer to move toward the average position of local flock mates).  The global, resulting picture are the mesmerizing patterns of abstract beauty that we all have seen at least once in the sky. Similar behavior have been studied for insects (swarming), quadrupeds (herding), fishes (schooling), but also for humans and robots in certain situations.  A second example is Twitter. Each user acts plainly: they tweet tiny messages, entirely self-interested or influenced by a small set of users they follow. But such micro posts, when multiplied by the mass of users, and channeled through the underlying labyrinthine network of followers, shape themselves into cultural shifts, global opinions, and even revolutions.\footnote{The crucial role of Internet and in particular of social networking services (Twitter in particular) during the uprisings of the Arab Spring has been largely acknowledged. These media have been used by insurgents to break isolation with the external world as well as to organize the internal revolution.} 

The phenomenon of complex systems whereby a simple conduct at the level of actors creates novel and coherent structures at a higher level is called \textit{emergence} \cite{H99}. Economist Jeffrey Goldstein provided a current definition of emergent phenomena, or emergents, in terms of the following properties \cite{G99}: (i) radical novelty: emergents are neither predictable from, deducible from, nor reducible to the micro-level components; (ii) coherence: emergents appear as integrated, unitary wholes that tend to maintain some sense of identity over time, in spite of the separation of the micro level components; (iii) macro level: the locus of emergent phenomena occurs at a global or macro level, in contrast to the micro-level locus of their components; (iv) dynamical: emergent phenomena are not pre-given wholes but arise as a complex system evolves over time; and (v) ostensive: emergents are recognized by showing themselves. Because of the nature of complex systems, each ostensive showing of emergent phenomena will be different to some degree from previous ones.  

These characteristics make emergence the ideal tool for the generative artist \cite{P11}. \textit{Generative art} is an art practice where the artist programs a system, which is set into motion with some degree of autonomy, contributing to or resulting in a completed work of art \cite{G03}. A defining feature of generative artworks is unpredictability: the generative artists cedes part of the control to the autonomous system in order to obtain an outcome that arouses surprise and emotion (radical novelty) and that shows itself different at every staging (ostensive). The generative artwork arises as a unitary whole as the autonomous system evolves in time (dynamical coherence), and the final artwork is at a higher granularity level with respect to the low level logic of the program and of the mechanics of the system (macro level). 

\begin{figure}[t]
\centering
\includegraphics[width=80mm]{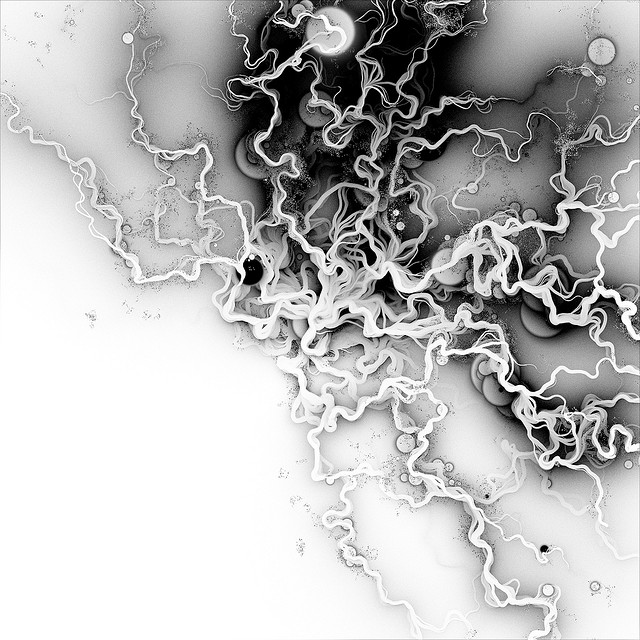}
\caption{Magnetic Ink by Robert Hodgin. The artist describes his artwork as follows: ``Magnetic Ink began as a tangent from the flocking studies I was working on at the time. The thinking was simple. What if the flocking birds rained down a fine mist of ink onto a sheet of virtual paper. At the same time, they have ribbons that hang from their feet and if they fly low enough, the ribbon will drag on the paper and erase the ink.''  \cite{RH14}}
\label{ink}
\end{figure}

Renowned exponents of the generative art movement include, to cite a few: Keith Peters, Jared Tarbell, Robert Hodgin, Marius Watz, Casey Reas, Paul Prudence, and Matt Pearson. To pick just one instance, Figure \ref{ink} shows Magnetic Ink, by Robert Hodgin. The intimate link between complex systems and generative art is witnessed by generative artist and complex science expert Philip Galanter, also curator of COMPLEXITY (Samuel Dorsky Museum Of Art, New Paltz, NY, USA, 2002), the first major fine art exhibition focusing on complex systems and emergence \cite{G03}:

\begin{quote}
``...the bulk of those working on the cutting edge of generative art are working with systems that combine order and disorder. These artists are exploring many of the same systems that are the very meat of complexity science. Examples include genetic algorithms, swarming behavior, parallel computational agents, neural networks, cellular automata, L-systems, chaos, dynamical mechanics, fractals, a-life, reaction-diffusion systems, emergent behavior, and all manner of complex adaptive systems.'' 
\end{quote}

\begin{figure}[t]
\centering
\includegraphics[width=120mm]{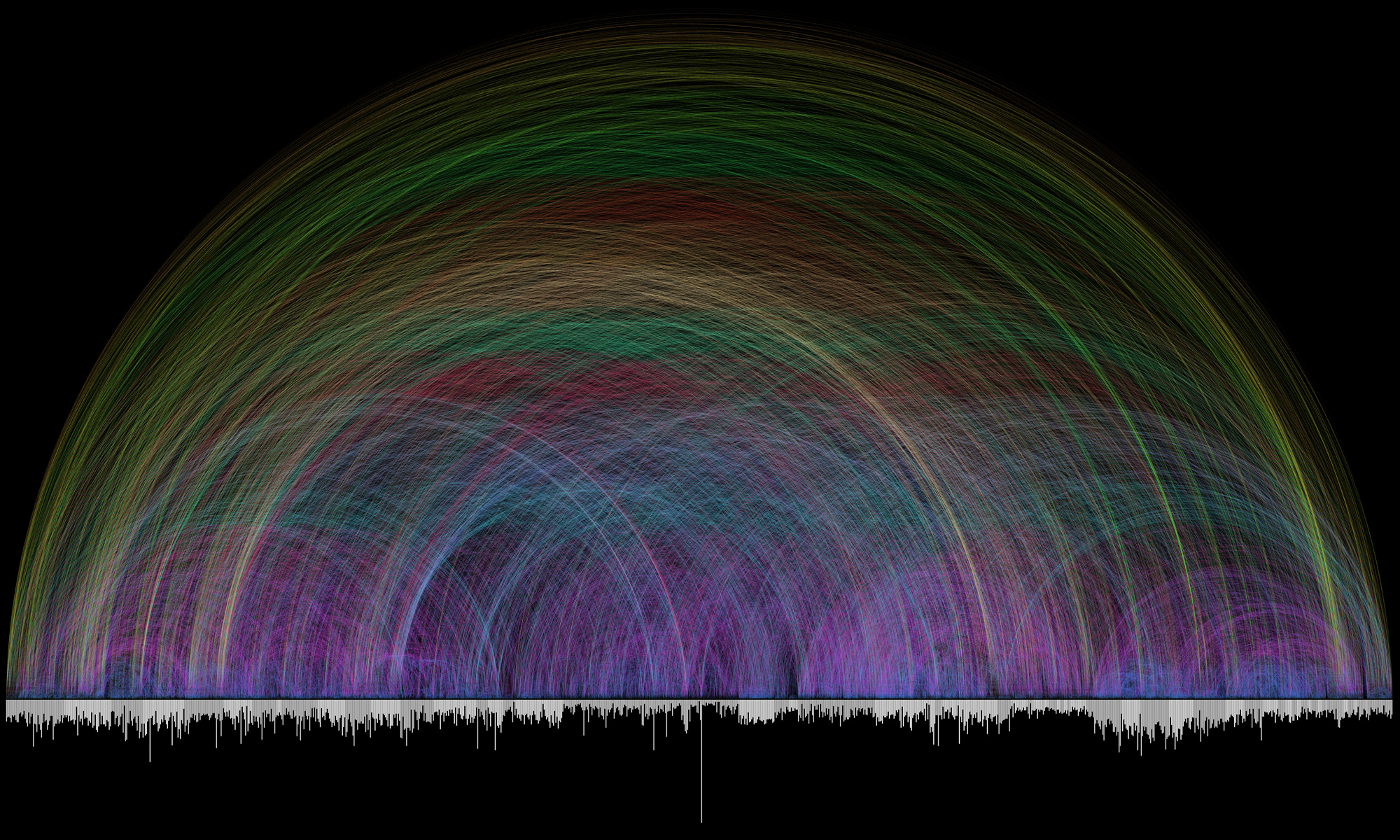}
\caption{Bible cross-references by Chris Harrison in collaboration with Lutheran Pastor Christoph R\"{o}mhild. This is how Harrison describes his artful visualization: ``The bar graph that runs along the bottom represents all of the chapters in the Bible. Books alternate in color between white and light gray. The length of each bar denotes the number of verses in the chapter. Each of the 63,779 cross references found in the Bible is depicted by a single arc - the color corresponds to the distance between the two chapters, creating a rainbow-like effect.'' \cite{CH14}}
\label{bible}
\end{figure}

But the contribution of complex systems to beauty and art overwhelms the generative art movement. Complex systems are ubiquitous; in particular, their most immediate and tangible manifestations, \textit{complex networks}, are at the focus of a philosophical, cultural and artistic chance of our highly interrelated and interdependent postmodern society. Rhizomatic structures offer a new model for knowledge and society aiming at acknowledging decentralization, autonomy, flexibility, creativity, diversity, collaboration, altruism and, ultimately, democracy \cite{DG80,L11,LHN05}. Networks match and sustain the proliferation of information typical of the postmodern condition, the co-existence of a multiplicity of heterogeneous discourses, instead of a simple, central discourse that unifies all forms of knowledge: ``Those who have a nostalgia for a unifying metanarrative - a dream central to the history of Western metaphysics - experience the postmodern condition as fragmented, full of anarchy and therefore ultimately meaningless. It leaves them with a feeling of vertigo. On the other hand, those who embrace postmodernism find it challenging, exciting and full of uncharted spaces. It fills them with a sense of adventure'' \cite{L84,C98}.  

\begin{figure}[t]
\centering
\includegraphics[width=100mm]{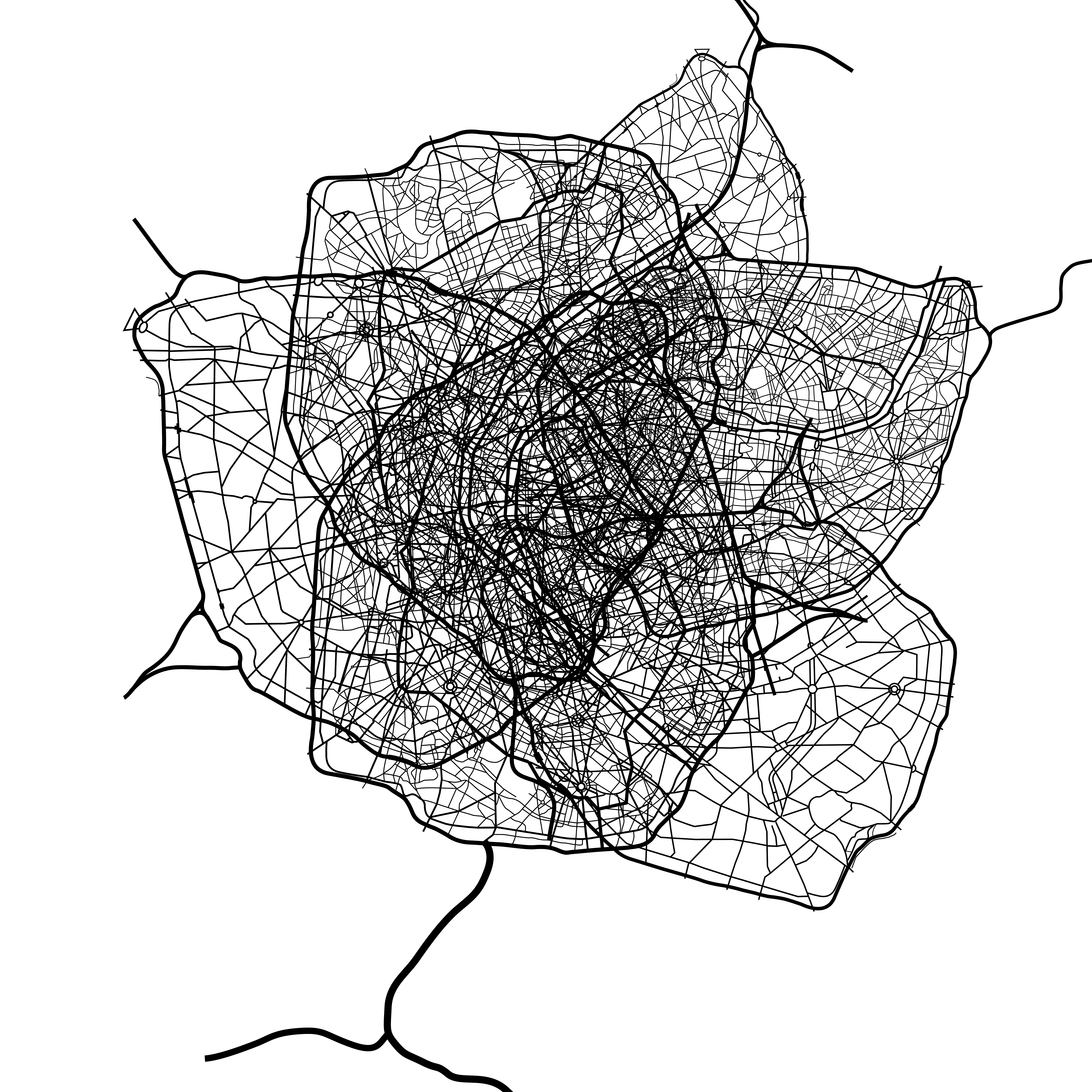}
\caption{ComplexCity Paris by Lee Jang Sub. The artist describes the project as follows: ``This project is an exploration to find a concealed aesthetic by using the pattern formed by the roads of the city which have been growing and evolving randomly through time, thus composing the complex configuration we experience today. I perceive the city's patterns as living creatures that I recompose to form an urban image.'' \cite{LJS14} }
\label{city}
\end{figure}

\begin{figure}[t]
\centering
\includegraphics[width=120mm]{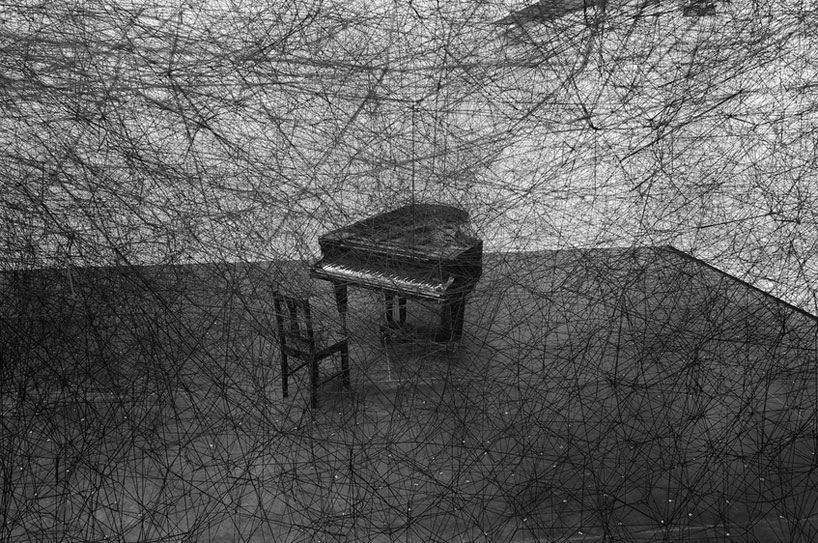}
\caption{In Silence, 2011, by Chiharu Shiota (photograph by Sunhi Mang). Material: Burnt grand piano, black wool. The artwork, featuring an abandoned, charred piano concert concealed beneath a complex network of interwoven yarn, is one of the best known installations of the artist \cite{CS14}.}
\label{piano}
\end{figure}

Such a perspective shift could not go unnoticed to the artist; a recognized function of art is to sense the times we are living and interpret them as a form of beauty, so that it can nurture our souls and caress our psychological frailties \cite{deBA13}.  Philosophers Gilles Deleuze and F\'{e}lix Guattari early envisaged the concept of network as an artwork, and more general as a cultural meme \cite{DG80}: ``the rhizome (...) can be torn, reversed, adapted to any kind of mounting, reworked by an individual, group, or social formation. It can be drawn on a wall, conceived of as a work of art, constructed as a political action or as a meditation.''  Manuel Lima, a creative mind and leading voice in information visualization, observes that ``complex networks are not just omnipresent, they are also intriguing, stimulating, and extremely alluring structures. Networks are not just the center of a scientific revolution; they are also contributing to a considerable shift in our conception of society, culture, and art, expressing a new sense of beauty.'' \cite{L11}  Lima is founder of \url{VisualComplexity.com} - a unified resource space for anyone interested in the visualization of complex networks. It showcases hundreds of beautifully visualized real complex networks, most of which are definitely artworks of reality. I opt for two absorbing examples: Bible cross-references, by Chris Harrison, depicted in Figure \ref{bible}, and ComplexCity, by Lee Jang Sub, illustrated in Figure \ref{city}.

In his captivating book \textit{Visual Complexity} \cite{L11}, moreover, Lima introduces the term \textit{Networkism} to identify a small but growing artistic trend, characterized by the portrayal of figurative graph structures of network topologies revealing convoluted patterns of nodes and links. Differently from network visualizations, which are based on a real dataset, the works produced by these artists, mainly paintings and sculptures, are fictitious. 
The influence of networkism is clearly visible in the works of Sharon Molloy, Emma McNally, Janice Caswell, Tomas Saraceno, Chiharu Shiota, Dalibor Nikolic, Akiko Ikeuchi, Ranjani Shettar, and Monika Grzymala, to cite a few, where imaginary landscapes of interconnected entities are the prevailing theme. See \url{networkism.org} for a digital portrayal of the artworks of these artists. An installation by Chiharu Shiota is pictured in Figure \ref{piano}. To gracefully end this section, this is how Sharon Molloy describes her work:

\begin{quotation}
My quest is to reveal how everything is interconnected. From the atom to the cell, to the body and beyond into society and the cosmos, there are underlying processes, structures and rhythms that are mirrored all around and permeate reality. (...) Ultimately I am trying to present a view of reality that reflects our changing times. This work embraces the multiple, the network, the paradoxical and the idea that even the smallest gesture or event has significance, and the power to change everything.
\end{quotation}

\section{Coda}
 
I have proposed the idea of \textit{artification} of science \cite{N12} and have exemplified the concept with the aid of complex systems and networks. I hope my endeavor is worthy and inspiring.
To my assessment, the benefits of a cross-pollination between science and art are several and include:

\begin{enumerate}[(a)]

\item Nonlinear approaches to the familiar increase your creativity and originality, two indispensable aspects of good research.   

\item New interesting problems arise, for instance: What is a suitable measure of complexity in aesthetics? Traditional complexity and information measures adopted in information theory, like Kolmogorov complexity and Shannon entropy do not work well here, since they equate randomness with maximal complexity and maximal information, while aesthetics considers randomness as interesting as boredom.

\item Your research tastes more interdisciplinary. In policy discourse interdisciplinarity is often perceived as a mark of good research - more successful in achieving breakthroughs and relevant outcomes \cite{RM10}. Moreover, speaking to different communities, interdisciplinary papers enjoy a larger window of visibility and a higher chance of sharing, and hence might accrue a greater impact measured in terms of alternative metrics, or altmetrics, which include the number of times a paper has been downloaded, tweeted, liked, covered by the media or blogs, cited on Wikipedia or bookmarked online \cite{K13}.

\item Your classes have a more stimulating flavor and attract more interested students. To my experience, students have a less specialized, more flexible mind and they are naturally inclined to appreciate interdisciplinary arguments. 
\end{enumerate}

\bibliographystyle{plain}

\end{document}